\documentstyle[12pt,agums]{article}

\lefthead{AFRAIMOVICH ET AL.}
\righthead{GPS PERFORMANCE DURING MAGNETIC STORMS}
\received{}
\revised{}
\accepted{}
\paperid{}
\authoraddr{E.~L. Afraimovich, V.~V. Demyanov, T.~N. Kondakova,
Institute of Solar-Terrestrial Physics SD RAS, p.~o.~box~4026, 
Irkutsk, 664033, Russia, fax: +7 3952 462557; e-mail:~afra@iszf.irk.ru}
\begin{document}
\title{Degradation of GPS performance
in geomagnetically disturbed conditions}
\author{E.~L. Afraimovich, V.~V. Demyanov, T.~N. Kondakova\\
Institute of Solar-Terrestrial Physics SD RAS,\\
p.~o.~box~4026, Irkutsk, 664033, Russia\\
fax: +7 3952 462557; e-mail:~afra@iszf.irk.ru}
\date{}
\begin{abstract}
The GPS performance is impaired in
conditions of geomagnetic distrubances. The rms error of
positioning accuracy increases in the case where two-frequency
GPS receivers of three main types (Ashtech, Trimble, and AOA) are
in operation. For Ashtech receivers (unlike AOA and Trimble)
there is also a clear correlation between the slip density of the
one- and two-frequency modes of positioning and the level of
geomagnetic disturbance.
\end{abstract}
\begin{article}
\section{Introduction}
\label{sec:intro}
The satellite navigation GPS system
[{\it Hofmann-Wellenhof et al.}, 1992] has become a powerful factor
of scientific and technological progress worldwide, and enjoys
wide use in a great variety of human activity. In this
connection, much attention is given to continuous perfection of
the GPS system and to the widening of the scope of its
application for solving the navigation problems themselves, as
well as for developing higher-precision systems for time and
accuracy determinations. Even greater capabilities are expected
in the near future through the combined use of the GPS with a
similar Russian system GLONASS [{\it Kharisov et al.}, 1998].

The performance of modern global satellite radio navigation
systems that utilize the "Earth-Space" radio wave
propagation channel is limited considerably by the influence of
the geospace environment. Furthermore, the main contribution
comes from systematic ionospheric effects of radio wave
propagation: the group and phase delay, the frequency Doppler
shift, and the rotation of the plane of polarization (Faraday
effect). In many instances the degree of manifestation of the
above effects has only a weak dependence on the local
distribution of electronic density in the ionosphere but is
directly correlated with the value of total electron content
(TEC) along the radio signal propagation path
[{\it Goodman and Aarons}, 1990].

In undisturbed geospace conditions the main contribution to the
formation of the above-mentioned ionospheric effects is
made by the regular TEC component. It undergoes periodic regular
variations (seasonal-diurnal, latitudinal, and longitudinal) and
is relatively accurately predictable. A variety of TEC models
have been developed to date, which are intended to cancel out the
ionospheric influence on the performance of the modern GLONASS
and GPS in geomagnetically quiet and weakly disturbed
conditions [{\it Afraimovich et al}., 2000a; {\it Klobuchar},
1986].

The situation with geomagnetically disturbed geospace is more
complicated. The irregular TEC component makes a substantial
contribution in this case. The amplitude of random TEC variations
with a period from a few minutes to several hours in conditions
of geomagnetic disturbances can make up as much as 50\% of the
background TEC value [{\it Basu et al}., 1988;
{\it Bhattacharrya et al}., 2000; {\it Ho et al}., 1996;
{\it Shaer et al}., 1997; {\it Warnart}, 1995].
Furthermore, the amplitude
and phase fluctuation range of signals from navigation satellites
(NS) at the reception point can exceed the designed level
corresponding to the uninterrupted operation of GPS receivers.
This leads to the degradation of the determination accuracy of a
current location (CL) of stationary and mobile users of GPS.
Furthermore, there might occur a break-down in tracking the NS
signal in phase (code) one of the working frequencies and, hence,
a failure in the determination of the coordinates in the one- or
two-frequency mode.

The study of deep, fast variations in TEC caused by a strong
scattering of satellite signals from intense small-scale
irregularities of the ionospheric $F2$-layer at equatorial and
polar latitudes has a special place among ionospheric
investigations based on using satellite (including GPS) signals
[{\it Aarons}, 1982; {\it Basu et al.}, 1988;
{\it Aarons et al.}, 1996, 1997; {\it Pi et al.}, 1997; {\it
Aarons and Lin}, 1999].

Recent years saw extensive studies of phase fluctuations and
phase slips of range measurements using GPS in conditions of
geomagnetic disturbances [{\it Afraimovich et al}., 2000; 2000a;
2001; 2001a; 2002; 2002a; {\it Bhattacharrya et al}., 2000; {\it
Skone and Jong}, 2000; {\it Coster et al.}, 2001; {\it Shan et
al}., 2002]. From the point of view of the GPS user, however, of
significantly greater interest are the investigations into the
influence of geomagnetic disturbances on the performance of GPS
as a positioning system.

The objective of this paper is to estimate the CL determination
accuracy and the failure density in determining the location in
the one-frequency and two-frequency modes in conditions of
geomagnetic disturbances for GPS receivers installed at permanent
mid-latitude stations forming part of the global GPS network.

\section{Experimental setup and data processing technique}
\label{SPE-sect-2}
In conducting the experiment in order to investigate the GPS
performance in conditions of geomagnetic disturbances, we
intended

- to estimate the CL determination accuracy of the GPS user in
conditions of geomagnetic disturbances as compared to undisturbed
conditions; and

- to establish the fact of presence (absence) of slips in the
determination of the user's location in the one-frequency mode of
standard accuracy and in the mode of two-frequency positioning in
conditions of geomagnetic disturbances, and
to obtain a numerical estimate of the slip density.

We used, as input experimental data, RINEX-files [{\it Gurtner},
1993] available through the Internet at
http://lox.ucsd.cdu/cgi-bin/allCoords.cgi. Of them, we considered
the data from the set of stations forming part of the global GPS
network (GPS stations) in the mid-latitude region (latitudes B -
$35...50^\circ$, longitudes L -$120...90^\circ$) in North America where the
distribution density of stations is the largest (the location map
of GPS stations is presented in Fig.1). We did not consider the
equatorial and polar regions.

Three days were used in the experiment: one quiet day (July 12,
2000, day number in the year 194), and two magnetically disturbed
days (July 15 and August 12, 2000, day numbers 197 and 225,
respectively) when strong magnetic storms occurred.

It is pointed out in [{\it Afraimovich et al}., 2000; 2000a;
2001; 2001a; 2002; 2002a] that geomagnetically disturbed
conditions show a different response of two-frequency receivers
manufactured by different firms. Thus, the lowest phase slip
density of range measurements at the working frequencies $f_1$ and
$f_2$ was observed for Ashtech receivers, a moderate density for
Trimble, and the highest density for AOA. In this connection, our
experiment was conducted with regards to GPS receivers of these
types. Table 1 presents general information about the experiment:
date; maximum level of geomagnetic disturbance characterized by
the maximum value of the geomagnetic index $-Dst_{max}$, nT, for 24
hours; GPS station name and location, and the name of the
receiver installed at a given station.

The goal of the first stage of GPS data processing in this paper
is to reconstruct the current location on the basis of available
RINEX files using software product TEQC posted by its developers
on the Internet at:
${\ http://unavco.ucav.edu/data-support/software/teqs/tewc.html}$
which we updated to ease the experiment.

In order to carry out investigations into the above-mentioned
issues, we developed software package "Navigator" to perform the
following functions:

- automated start of software product TECQ in the package mode by
setting up a current command line of the form: $teqc.exe +qc = st
YYmmddhh/MMSS + dh  \Delta{t} - nav sitesddd0.YYn sitesddd0.YYo$, where
$teqc.exe$ is the executed file of software product TEQC; $qc$, $st$,
and $dh$ are the control options in the command line; $YY$, $mm$, $dd$,
$MM$, and $SS$ are the year, month, day, hour, minute, and second for
which the coordinates of the object are calculated with a current
start of TEQC; $\Delta{t}$  is the time step of calculation of the
coordinates on the interval of observation (in fractions of an
hour); and $sites$ is the abridged name of the GPS station, to
which the RINEX-files of format "o" (observation) and "n"
(navigation) refer.

- setting up of the output data file in the format:

$t_{i}; X_{i}; Y_{i}; Z_{i}; L_{i}; B_{i}; H_{i}; \Delta{X}_{i}
;
\Delta{Y}_{i}; \Delta{Z}_{i}; \Delta{L}_{i}; \Delta{B}_{i}; \Delta{H}_{i}$,
 where $t_{i}$ is a current time (in hours and in fractions of
an hour); $X_{i}, Y_{i}$, and  $Z_{i}$ are current rectangular geocentric
coordinates (in meters); $L_{i}, B_{i}$, and $H_{i}$ are current geocentric
coordinates; and   $\Delta{X}_{i};  \Delta{Y}_{i};  \Delta{Z}_{i};
\Delta{L}_{i};
\Delta{B}_{i}; \Delta{H}_{i}$ are absolute current errors of
determination of the corresponding rectangular and geocentric coordinates.

- detecting the slips of the one- and two-frequency mode of
positioning, and determining the corresponding slip density;

- estimating the current and daily mean spherical standard
deviations of CL determination for the GPS station.

The software package is controlled through the interactive file
in which the following data are specified:

- name of GPS station sites;
- form of the command line (if necessary, the form of the command
line can be changed);
     - date (day, month, day);
- start and end time of the period of observation (in hours,
minutes and seconds);
- size of the time step of calculation of the current coordinates
($\Delta{t}$: from 0.018 to 24 hours);
- direction of count time increment -- toward increasing ($+dh$) or
decreasing ($-dh$), respectively;
- ON- (OFF-) option of the block of analyzing the slip density
and the stands deviations of CL determination.

As a result of the first stage of processing, for each GPS
station from Table 1 we reconstructed for three days (194, 197,
and 225) of the year 2000 the diurnal series of rectangular
geocentric coordinates $X_{i}; Y_{i}; Z_{i}$; at time steps of 1.2 minute
(the minimum possible steps when using the TEQC package), and the
corresponding absolute errors:

$\Delta$$x$=$x_{i}-x_{0}$; $\Delta$$y$=$y_{i}-y_{0}$; $\Delta$$z$=$z_{i}-z_{0}$,

where $x_{0}$, $y_{0}$, $z_{0}$ are the known coordinates of the GPS station,
and $i$ is the time count number.

The second stage of data processing involved estimating the
current CL determination accuracy that is equal to the value of
the spherical standard deviation (SD) in the determination of the
coordinates $\sigma(t_{i})$, m:

$\sigma(t_{i})=(\sigma_{xi}+\sigma_{yi}+\sigma_{zi})^{0{.}5}$,

where $\sigma_{xi},\sigma_{yi},\sigma_{zi}$ are the SD in the determination 
of the corresponding coordinate in a rectangular geocentric coordinate system. 
Current values of $\sigma(t_{i})$ were calculated for each 3.6-minute interval 
over the length of the entire diurnal series.

In addition, for a generalized estimation of the determination
accuracy of the coordinate of each station in particular
geomagnetic conditions, we calculated the daily mean values of
the spherical SD $\overline{\sigma}$, m. Results of this processing are plotted as
$\sigma{(t)}$ in Figs.2,3,4,5,6,7 are presented as histograms for $\overline{\sigma}$
(Fig.8).

The third stage of data processing included estimating the slip
density in the determination of the CL for receivers of three
types (Ashtech, AOA, and Trimble) in the one- and two-frequency
modes. In the analysis of the slip intensity in the determination
of the coordinates in the one-frequency mode, the slip was
considered to mean the event implying that the following
condition is satisfied:

             $\sigma{(t)}\leq\Pi=500$ m

We deduced empirically the value of the unacceptable error
threshold  $\Pi=500$ m, based on the fact that the best
determination accuracy of CL is required for most practical
purposes [{\it Skone and Jong}, 2000].

By the slip in the mode of two-frequency determination of the
coordinates was meant the event implying that at a current
instant of time, more than 30$\%$ of all observed NS of the
constellation do not provide tracking at the auxiliary working
frequency $f_2$.

As the characteristic of the slip intensity of the one- and
two-frequency mode of coordinate determination, we calculated the
current slip densities of the one- and two-frequency modes ($N_1$
and $N_{1,2}$, respectively). The current slip density $N_1$ and $N_{1,2}$ was
determined as the total number of slips on each 0.5-hour
interval. Results of this processing are presented as plots of
the diurnal slip density distribution for the one-frequency mode
(Fig.10,11,12), and for the two-frequency mode (Fig.13,14,15).

Furthermore, we also determined the total diurnal slip density Ns
for the one- and two-frequency modes, for each station and for
each of the days listed in Table 1. Results of this processing
are represented by the histograms in Figs.9 and 16, respectively.

\section{Results and discussion}
\label{SPE-sect-3}

\subsection{The magnetically quiet day of July 12, 2000}
\label{SPE-sect-3.1}

Fig.2 (panels b,c,d) and Fig.3 (b,c) plot the time dependencies
of the spherical standard deviations $\sigma{(t)}$ in the CL determination
on the magnetically quiet day of July 12, 2000, obtained at three
different sites: Kew1, Stb1, and Chb1 with the Ashtech receivers,
and at sites Upsa (Trimble) and Algo (AOA). Table 2 presents the
corresponding daily mean values of $\overline{\sigma}$. On some of the panels
where the maximum values of standard deviations far exceeded the
daily mean values, these values of the maximum diurnal standard
deviation  $\sigma_{max}$ are included in Table 2 and removed from the
figures to ease the reading. A more general picture of the daily
mean values of standard deviations for all types of receivers for
the days analyzed in this paper is provided by the Fig.8
histograms. In this figure, each type of receiver under
investigation is represented by a set of three stations. Ashtech
is represented on panel a) by the sites Kew1, Stb1, and Chb1; AOA
is represented on panel b) by the sites Algo, Gode, and Usno; and
Trimble is represented on panel c) by the sites Tung, Dyer, and
Upsa.

It is evident from the figures and histograms presented here that
in magnetically quiet conditions the CL determination accuracy
that is provided by the Ashtech and AOA receivers differs only
slightly (for Ashtech $\overline{\sigma}\leq46$ m, and for AOA
$\overline{\sigma}\leq33$ m). Single abrupt changes of the
current values of $\sigma{(t_i)}$ over the course of 24 hours
occur for the three types of receivers (Table 2). Furthermore,
their maximum values do not differ greatly $\sigma_{max}\leq249$
m for Ashtech; $\sigma_{max}\leq256$ m for AOA, and
$\sigma_{max}\leq243$ m for Trimble.

As regards the number of slips in the coordinate determination in
the one-frequency mode under quiet geomagnetic conditions, and
Ashtech and AOA receivers also showed about the same level. Thus,
no more than three slips of the one-frequency mode and no more
than two slips were observed for 24 hours, respectively for
Ashtech (site Kew1) and AOA (site Gode). For the Trimble
receiver, this figure was worse and reached ten slips for 24
hours (site Tung). On the other hand, the three types of
receivers at some of the sites did not show any slips of the
one-frequency mode for 24 hours (see Fig.9). All observed slips
occurred before 12 UT.

The performance of the Trimble receivers in the two-frequency
mode of coordinate determination in the magnetically quiet
situation was also worse when compared with Ashtech and AOA.
Thus, for the receivers of the two last types, no slip of the
two-frequency mode was recorded for all the sites under
consideration (see Fig.13).

\subsection{The magnetic storms of July 15 and August 12, 2000}
\label{SPE-sect-3.2}

In general, the period of geomagnetic disturbances showed a
degradation of the accuracy and quality of GPS performance for
the three types of receivers.

During the July 15 magnetic storm there was an increase of the
value of the daily mean spherical standard deviation in the CL
determination $\overline{\sigma}$, m. The corresponding maximum
values were as follows: 76 m for the Ashtech receiver (site
Kew1); 39 m for AOA (site Algo), and 120 m for Trimble (site
Upsa). Thus the value of $\overline{\sigma}$ increased (compared
to geomagnetically quiet conditions) by a factor of 1.6-1.7 for
Ashtech, and by a factor of 1.1-1.2 for AOA and Trimble (for the
sites Algo and Upsa, respectively). The estimates made here and
Figs.3,6 and 8 suggest also the conclusion that the Ashtech
receivers were the most sensitive to a change of the geomagnetic
situation. Values of the spherical SD in the CL determination
using the AOA and Trimble receivers do not increase as
significantly as above and can, in some cases, also be lower than
those in geomagnetically quiet conditions (Fig.8 (a-c)).

A similar picture is also observed during the magnetic storm of
August 12, 2000; however, the effect is more fully manifested.
There was an unambiguous increase of the values of $\overline{\sigma}$ for the
three types of receivers and for all sites when compared with
magnetically quiet conditions. Values of $\overline{\sigma}$ increased by a
factor of 1.6-2.6, 1.3-2.1 and 1.02-1.4 for Ashtech, AOA and
Trimble (Fig.4 (b, c, d); 7 (b, c), and 8).

Table 2 presents the maximum values of single abrupt changes of
current values of the spherical SD  in  the  CL  determination $\sigma{(t_i)}$
for the 24 hours that are analyzed. For the Ashtech and Trimble
receivers, there is also a consistent tendency of the values of
these abrupt changes to increase in geomagnetically disturbed
conditions. Figs.3 (b, c); 4 (b, c, d) induce us also to suggest that for
Ashtech the intervals with the largest number of abrupt changes
of current values of $\sigma{(t_i)}$ correspond to the time period of the
highest level of geomagnetic field disturbance. However, to
confirm this assumption requires an extensive set of
observations.

On the basis of studying the slip density and localization of the
one- and two-frequency modes in the CL determination, the
following features were established.

Firstly, we identified an unambiguous tendency of the slip
density of the one-frequency mode to increase in magnetic storm
conditions for the Ashtech and Trimble receivers. Compared to
magnetically quiet conditions, the slip density of the
one-frequency mode increased by a factor of 2.5 and 1.4-8 for
Ashtech and Trimble, respectively (Fig.9, Figs.10 (b, d); 11 (b, d);
12 (b, d)).

Secondly, for the Ashtech receiver the localization of slips of
the one-frequency mode along the time axis corresponds to the
time period of maximum geomagnetic field disturbance (Figs.11b;
12b).
The sole exception was the Ashtech receiver at the site Chb1. No
slips of the one-frequency mode was recorded for this site for
any one of the days under consideration.

Thirdly, we detected a clearly pronounced tendency of the slip
density of the two-frequency mode in the CL determination to
increase for the Ashtech receivers in geomagnetically disturbed
conditions. Thus, while no slips were observed for the
magnetically quiet day of July 12, at periods of geomagnetic
disturbances their diurnal number varied from 40 (site Stb1,
August 12) to 120 (site Kew1, July 15) -- Fig.16.

Besides, there was a clear localization of the two-frequency mode
for the time interval corresponding to the maximum level of
geomagnetic field disturbance (Figs.14 (a, b) and 15 (a, b)).
A similar, but less
clearly pronounced, tendency of the slip density of the
two-frequency mode to increase was also identified for the
Trimble receivers (Fig.16b). We recorded no slips of the
two-frequency mode in magnetic storm conditions for the AOA
receivers in all cases that were considered.

Thus our investigation has shown that geomagnetic disturbances of
geospace are accompanied by a degradation of the accuracy and
quality of GPS performance. During geomagnetic disturbances
of geospace the Earth's ionosphere becomes essentially
inhomogeneous. The dramatic irregular geospace variations during
geomagnetic disturbances are associated with the generation in
the ionosphere of a broad spectrum of irregularities giving rise
to the scattering and defocusing of the satellite signal.

As a result, receivers can observe significant signal amplitude
fluctuations capable of causing the break-down of its tracking.
Besides, the presence of irregular electron density
irregularities along the radio signal propagation path is
responsible for the proportionate fluctuations of the group and
phase delay [{\it Yakovlev}, 1996] which degrade the accuracy of
CL determination. Abrupt fluctuations of the phase delay can also
cause the breakdown of NS signal tracking if the frequency
Doppler shift of the received signal exceeds the frequency
bandwidth of the tracking contour in phase.

As has been shown above, slips of the two-frequency mode of
coordinate determination are a more common phenomenon compared
with slips of the one-frequency mode. The main reason for this
might be the fact that the signal level at the auxiliary (closed)
frequency $f_2$ is lower than that at the fundamental frequency
$f_1$. The received power, with the elevation angle of the ray to
the NS of $45^\circ$, is 159 Db/W at the frequency $f_1$, and 166
Db/W at the closed frequency $f_2$ [{\it Interface Control
Document}; Langley, 1998]. As far as the differences between the
response of the particular types of GPS receivers to geomagnetic
disturbances of geospace are concerned, this issue still remains
open and is beyond the scope of this paper. To elaborate on this
question requires analyzing the design and functional features of
receivers manufactured by different firms.

\section{Conclusion}
\label{SPE-sect-4}

Main results of this study are as follows:
\begin{enumerate}
\item In geomagnetically disturbed conditions of geospace the
accuracy and quality of GPS performance is impaired.

\item Unlike geomagnetically quite conditions, the magnetic storm
conditions are accompanied by an increase of the spherical
standard deviation in the location determination for all types of
GPS receivers. Furthermore, there was a maximum increase of the
values of $\overline{\sigma}$ by a factor of 2.6, 2.1 and 1.4 for
the Ashtech, AOA and Trimble receivers, respectively.

\item In magnetic storm conditions there is an increase of the
number of slips of the one-frequency mode of coordinate
determination. The slip density increased by a factor of 2.5, 8
and 12 for the Ashtech, Trimble and AOA receivers, respectively
(site Usno).

\item We have identified an unambiguous tendency of the slip density
of the one-frequency mode to increase for the Ashtech and Trimble
GPS receivers, respectively. For the AOA receivers this tendency
was observed not in all cases that were considered.

\item The slip density in the two-frequency mode of coordinate
determination in magnetic storm conditions increases most
dramatically for the Ashtech receivers (from 0 to 120 slips). As
regards the Trimble receivers, there is a similar, but less
clearly pronounced, picture (the slip density increased by a
factor of 1.5-2, on average). No slips of the two-frequency mode
was detected for the AOA receivers in the cases under
consideration.

\end{enumerate}
\acknowledgments
The authors are grateful to G. V. Popov and V. G. Eselevich for their
encouraging interest in this study and active participation in discussions. 
The authors are also indebted to ~E.~A.~Kosogorov and ~O.~S.~Lesuta for 
preparing the input data. Thanks are also due V.~G.~Mikhalkovsky for his 
assistance in preparing the English version of the \TeX manuscript. 
This work was done with support from the Russian Foundation for Basic Research 
(grants 01-05-65374 and 00-05-72026) and from RFBR grant of leading 
scientific schools of the Russian Federation 00-15-98509.

{}
\end{article}
\end{document}